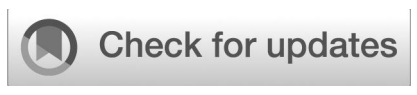

# Nationwide analysis of medicine search behaviour and system performance: a one-year evaluation of MediVerify in Sri Lanka (July 2024–July 2025) based on 1.49 million search queries

Praveen Charuka Athauda-Arachchi[1*], Pandula Mahesh Athauda-Arachchi[2*] and Rohini Fernandopulle[2,3]

[1]British School in Colombo, Sri Lanka, [2]Department of Pharmacology and Cardiology, Faculty of Medicine, General Sir John Kotelawala Defence University, Ratmalana, Colombo, Sri Lanka, [3]Faculty of Medicine, University of Colombo, Colombo, Sri Lanka

**Background:** National-scale digital health platforms generate large volumes of usage data that provide insights into population-level healthcare needs. MediVerify, Sri Lanka's national online medicines information platform, providing public access to the registry of all National Medicines Regulatory Authority (NMRA)-approved products, recorded over 1.49 million queries in its first year of operation. However, systematic analysis of such large-scale medicine search behaviour in low- and middle-income countries (LMICs) remains limited.

**Objectives:** (1) Characterise national medicine information-seeking behaviour using large-scale search logs; (2) identify mismatches between public demand and essential medicines policy; and (3) evaluate the technical performance and computational efficiency of a national digital health system.

**Methods:** We conducted a retrospective observational analysis of 1,497,304 anonymised search queries on MediVerify between July 2024 and July 2025. Queries were normalised and mapped to a registry of 11,933 approved medicines using a multi-step pipeline including fuzzy matching (Levenshtein distance, threshold $\leq 5$). Therapeutic categories were assigned using an Anatomical Therapeutic Chemical (ATC)-aligned classification. Query distributions, zero-query and zero-result patterns, and system performance metrics were analysed. Energy consumption was estimated from observed processing times.

**Results:** Vitamins/minerals (11.74%), antibiotics (10.57%), diabetes medications (7.02%), and antihypertensives (6.84%) dominated searches. The top 20 medicines comprised 37.3% of all queries. Approximately 40%–50% of registered medicines were never queried. Zero-result queries (~1%) revealed unmet information needs. A significant divergence was observed between frequently searched medicines and essential medicines list (EML) designations. System performance was robust, with a median latency of 8 ms and >99% query resolution. Estimated energy consumption was extremely low (~0.12 kWh per million queries).

OPEN ACCESS

*CORRESPONDENCE
Praveen Charuka Athauda-Arachchi
praveenathaudaarachchi@gmail.com
Pandula Mahesh Athauda-Arachchi
athaudaarachchipm@kdu.ac.lk

**Conclusions:** Large-scale medicine search data provide actionable insights into national healthcare information demand and policy alignment. MediVerify demonstrates that a national digital health platform can achieve high utilisation with minimal computational and environmental cost. Findings support integration of real-time usage analytics into pharmaceutical policy and highlight the potential of low-resource digital infrastructure in strengthening health systems in LMICs.

# 1 Introduction

Digital health infrastructure has become integral to national medicines governance, supporting transparency, public access, and regulatory efficiency. Sri Lanka's NMRA previously relied on a publicly accessible PDF-based medicines registry; however, limitations in scalability, update frequency, and searchability created operational inefficiencies. MediVerify was developed to address these gaps by providing a real-time, searchable digital registry of all NMRA-approved medicinal products. The initial evaluation of MediVerify's launch week documented over 21,700 searches (averaging ∼3,100 per day), with paracetamol and antibiotics dominating early query trends (1). This early success suggested considerable public demand for accessible medicines information (2). The platform also helps regulators and the public to understand registration status and to seek contact information of licence holders in times of product deficits.

Despite this promising start, long-term national search patterns had not been characterised. Understanding a full year of query data offers deeper insight into population-level information needs, therapeutic demand trends, and the alignment—or mismatch—between public queries and essential medicine priorities.

Similar platforms internationally have demonstrated that real-time query analytics can contribute to pharmacoepidemiology, shortage detection, and policy refinement. NHS Open Prescribing in England provides open access to national prescribing data and has been used to monitor antibiotic stewardship and identify regional anomalies (3). The U.S. FDA National Drug Code (NDC) directory serves as a comparable resource at national scale (4). The Health Sciences Authority (HSA) of Singapore maintains a publicly accessible medicines database used for regulatory verification and formulary decisions (5). However, published analyses of analogous platforms in LMICs remain sparse.

A 2022 systematic review of consumer health information-seeking behaviour in LMICs (Lagan et al., BMC Health Services Research) identified substantial unmet demand for accurate medicines information, particularly among populations with high non-communicable disease (NCD) burdens (6). A 2023 study by Moon et al. (Bulletin of the WHO) found that 30%–60% of the most frequently prescribed medicines across twelve LMICs were absent from the EML (7). A 2024 analysis of digital medicines verification platforms in sub-Saharan Africa (Nkosi et al., Digital Health) reported that low-resource text-based tools could generate actionable pharmacovigilance data at national scale (8). These studies collectively establish that national digital medicines platforms in LMICs can yield public health intelligence relevant to health systems strengthening.

Building on this context, our study provides the first comprehensive year-long analysis of MediVerify query behaviour in Sri Lanka. To our knowledge, this is the first large-scale analysis of national medicine search behaviour in an LMIC, and one of the first studies to quantify the computational and environmental efficiency of a national digital health search system. The objectives of this study are to: (1) characterise national medicine information-seeking behaviour using large-scale search data; (2) identify mismatches between public demand and essential medicines policy; and (3) evaluate the technical performance and efficiency of the MediVerify platform.

# 2 Methods

## 2.1 Research design

This study adopts a post-positivist epistemological paradigm and a quantitative research approach, operationalised through a hypothetico-deductive methodological framework (9). The research design is non-experimental, retrospective, and observational in nature, consistent with the analytical study type described by Hulley et al. (10) and the classification of observational study designs proposed by Thiese (11). This design was appropriate for analysis of pre-existing system-generated data without any intervention or manipulation of study variables.

## 2.2 Study design and data source

We conducted a retrospective observational analysis of all search queries submitted to the MediVerify platform from 11 July 2024 through 10 July 2025. Each query was time-stamped and anonymised (no user-identifying data were collected). In total, 1,497,304 search entries were recorded during the study period. The registry of approved medicines maintained by the NMRA (containing 11,933 product entries as of 2024) was used as the reference database for mapping queries.

## 2.3 Query processing and classification

All search strings underwent a multi-step normalisation pipeline: (i) Unicode normalisation to NFC form; (ii) case folding to lowercase; (iii) removal of punctuation, diacritics, and leading/trailing whitespace; and (iv) collapsing of internal whitespace to single spaces. The normalised query was then matched against the NMRA medicines database, which indexes both the approved International Non-proprietary Name (INN/generic name) and all registered brand names for each product, enabling brand-to-generic and generic-to-brand query resolution within a single lookup.

If an exact match (edit distance = 0) was found, the query was classified accordingly. If not, the platform's fuzzy matching engine applied the Levenshtein distance algorithm against all indexed medicine names. A match was accepted if the edit distance did not exceed a predefined threshold of five characters ($\leq 5$), selected empirically to capture common misspellings (e.g., transpositions, omissions, substitutions) while avoiding implausible matches. Results were ranked by ascending edit distance; ties were resolved by database registration frequency. Where a query matched multiple entries at equal edit distance, all results were returned to the user; for classification purposes, the highest-frequency entry was used as the primary match. Queries exceeding the threshold were recorded as zero-result queries.

Each query was assigned to a therapeutic category based on an ATC-aligned classification adjusted to reflect local prescribing patterns. Top 20 queried medicines were identified from percentage breakdowns converted to absolute counts (count = %/100 × 1,497,304). Zero-query medicines (never searched during the year) were identified by cross-referencing queried names against the full NMRA database. Medicines with ambiguous or duplicate name mappings were manually reviewed. Alignment with essential medicines was evaluated against the 2025/2022 WHO Model EML (16, 18) and the 2022 Sri Lanka National EML (17).

## 2.4 System performance analysis

For each query, the system-recorded latency (ms), edit distance, query string length, and number of results returned were analysed. Latency was summarised using descriptive statistics (mean, median, 90th, 95th, and 99th percentiles) to characterise typical vs. worst-case response times. The relationship between edit distance and latency was examined using Pearson's $r$.

To estimate the computational and energy footprint of the platform, we used the observed average processing time per query in conjunction with a conservative server power draw (~10 W for a small cloud instance or on-premises server) (12–14). Energy per query was calculated in joules and extrapolated to kilowatt-hours per million searches. Carbon footprint was estimated using a Sri Lanka grid electricity emission factor of 0.5–0.7 kg $CO_2$/kWh. An independent system-level analysis used hourly CPU utilisation over 8,640 observations, segmented by calendar month and four equal study quartiles (Q1–Q4). Because the distribution was markedly non-normal, non-parametric tests were applied throughout: global differences across months and quartiles were tested using the Kruskal–Wallis test; Friedman tests were applied to hourly median profiles; *post-hoc* pairwise Mann–Whitney U comparisons were performed with Holm correction for multiple testing.

# 3 Results

Results are presented in three sections corresponding to the three study objectives.

## 3.1 Objective 1: national medicine information-seeking behaviour

### 3.1.1 Distribution of searches by therapeutic category

Searches were widely distributed across therapeutic classes. Vitamins/minerals comprised 175,739 searches (11.74% of total), antibiotics 158,315 (10.57%), diabetes-related drugs 105,165 (7.02%), and antihypertensives 102,358 (6.84%). Together these four categories represented >36% of all searches. Other notable categories included analgesics, respiratory medications, lipid-lowering agents, antifungals, and antipsychotics (Figure 1).

### 3.1.2 Most searched individual medicines

The top 20 individual medicines accounted for approximately 37.3% of all queries (~558,000 searches), indicating a moderately long tail of less-frequent queries outside the top hits. The highest-ranked search term was rosuvastatin, a cholesterol-lowering medication (48,812 searches; ~3.26%). Paracetamol was a close second (45,967; 3.07%) and glimepiride (an oral hypoglycaemic) ranked third (45,668; 3.05%). Other leading medicines included montelukast (39,080; 2.61%), ondansetron (38,930; 2.60%), and vitamin E supplements (33,689; 2.25%).

### 3.1.3 Zero-query registered medicines

Approximately 40%–50% of all registered medicines were never queried during the study year. These were largely specialised or seldom-used products, including oncology drugs, biologic therapies, antidotes, paediatric-specific formulations, or hospital-restricted medications. The high fraction of zero-query products underscores the skewed distribution of public interest: a core subset of medicines attracts the majority of attention, while a long tail of products remains virtually unseen by users—a pattern consistent with healthcare utilisation data from comparable national platforms internationally.

### 3.1.4 High-frequency zero-result searches

Zero-result searches comprised approximately 1% of all queries. High-frequency zero-result terms included: queries for products not registered in Sri Lanka (e.g., unregistered emergency contraceptive brand names); mebendazole (an anthelmintic apparently absent from the NMRA registry during the study period, suggesting a regulatory gap or naming mismatch); brand-generic mismatches (users searching a brand name not indexed as such in the registry); and spelling variations of common drug names. A smaller subset involved non-medicinal products or general terms not covered by the medicines registry.

### 3.1.5 Interquartile variability of top 20 searched medicines

There was a highly significant difference in search frequencies across quarters (Friedman $\chi^2 = 43.32$, $p = 2.10 \times 10^{-9}$) (Figure 2). The effect size was large (Kendall's $W = 0.72$), indicating strong temporal variation in medicine search behaviour. *post-hoc* pairwise comparisons (Wilcoxon signed-rank, Bonferroni correction) demonstrated that search volumes in Q3 and Q4 were significantly lower than those in Q1 and Q2 ($p < 0.001$ for all comparisons). No statistically significant difference was observed between Q1 and Q2 (adjusted $p = 0.091$), suggesting a plateau of peak demand during the first half of the year. Q4 exhibited significantly lower search volumes compared to Q3 ($p = 0.0047$).

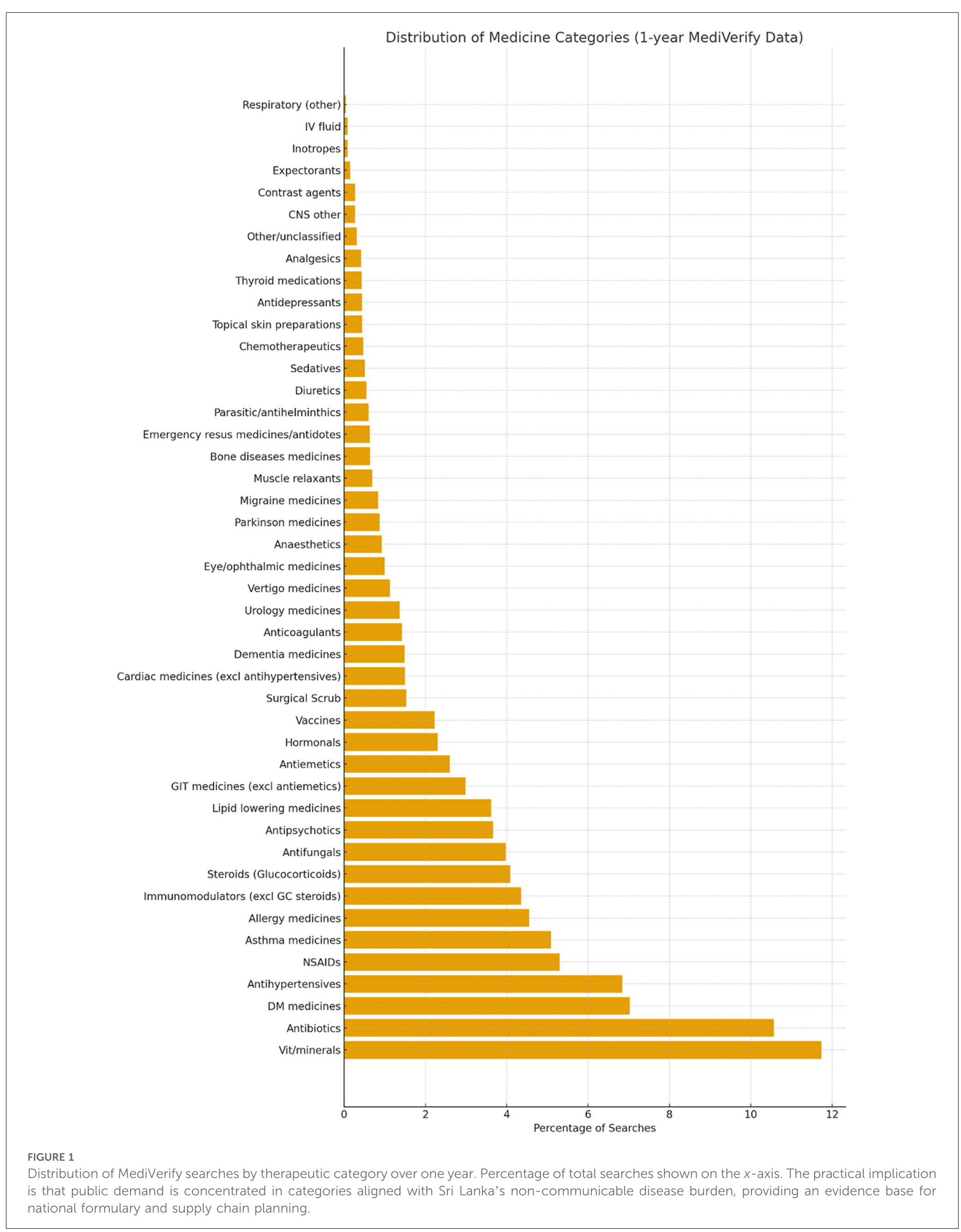

FIGURE 1
Distribution of MediVerify searches by therapeutic category over one year. Percentage of total searches shown on the $x$-axis. The practical implication is that public demand is concentrated in categories aligned with Sri Lanka's non-communicable disease burden, providing an evidence base for national formulary and supply chain planning.

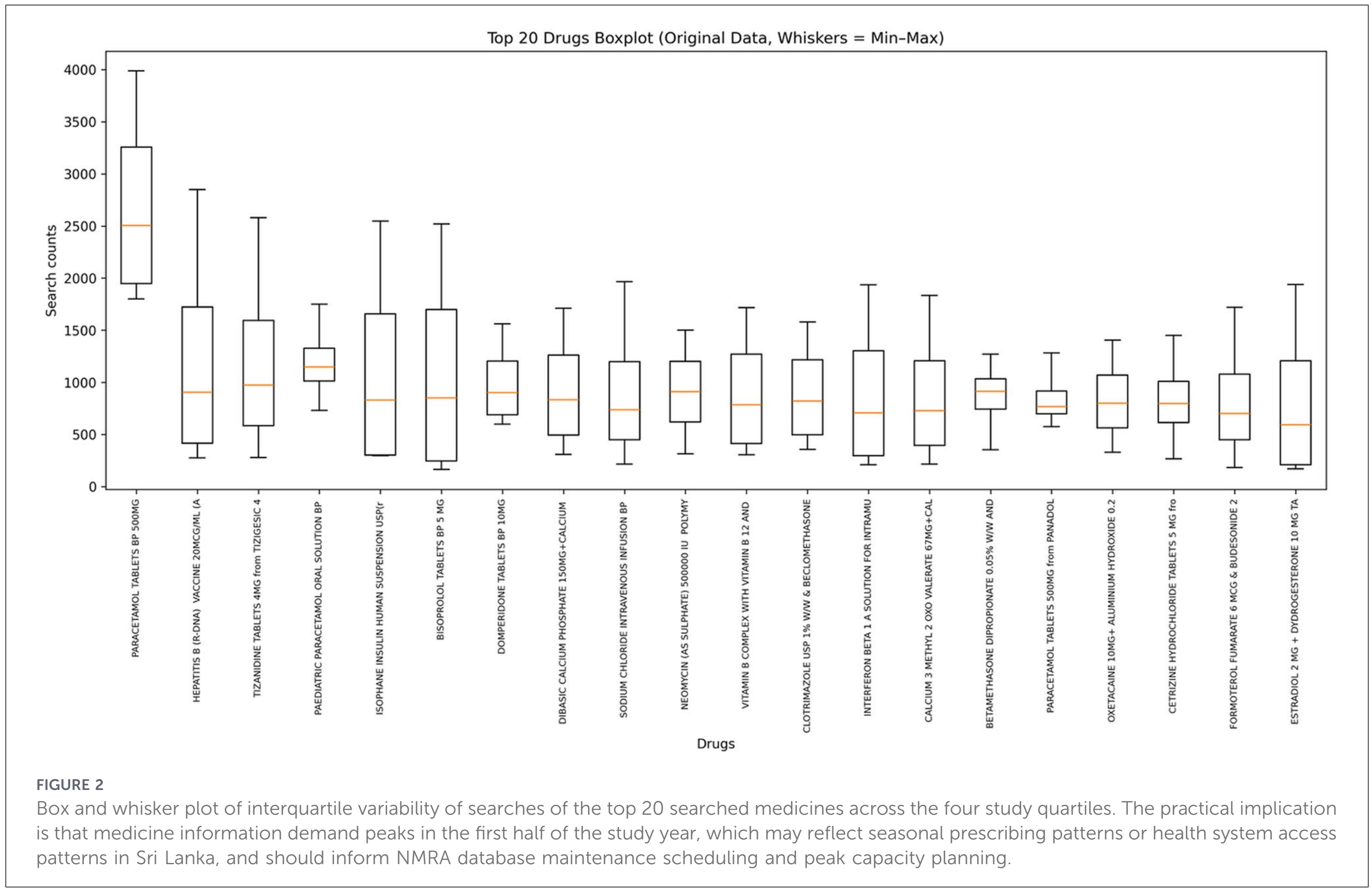

FIGURE 2
Box and whisker plot of interquartile variability of searches of the top 20 searched medicines across the four study quartiles. The practical implication is that medicine information demand peaks in the first half of the study year, which may reflect seasonal prescribing patterns or health system access patterns in Sri Lanka, and should inform NMRA database maintenance scheduling and peak capacity planning.

## 3.2 Objective 2: alignment with essential medicines policy

### 3.2.1 Essential medicines list cross-reference

Table 1 presents the top 20 searched medicines and their status on the 2025 WHO Model EML and the 2022 Sri Lanka National EML. Nine of the top 20 queried medicines (including rosuvastatin and glimepiride) were not listed on either EML. Nearly 70% of the top 20 were absent from one or both essential medicines lists, highlighting a significant divergence between public search interest and official essential medicine designations. This misalignment may reflect evolving disease burdens, prescribing practices, or public perceptions, and highlights areas where essential medicines frameworks may require reassessment.

## 3.3 Objective 3: technical performance and computational efficiency

### 3.3.1 Search accuracy and fuzzy matching performance

Approximately 85% of all searches were resolved by an exact string match (edit distance = 0). The remaining ∼15% required fuzzy matching (edit distance ≥1) due to misspellings or partial names. About 10% of queries had an edit distance of 1; ∼3% had an edit distance of 2; only ∼2% had an edit distance ≥3, confirming that substantial misspellings were relatively rare (Figure 3). The Pearson correlation coefficient between edit distance and latency was $r \approx 0.69$ ($p < 0.001$), indicating a moderate-to-strong positive relationship (Figure 4).

### 3.3.2 System latency

Across all queries, the mean system response time was approximately 42 ms. However, the distribution of latencies was highly right-skewed: the median latency was only ∼8 ms, meaning half of all searches returned results almost instantaneously (Figure 5). The 90th percentile latency was ∼45 ms; the 95th percentile was ∼100 ms; and the 99th percentile was ∼800 ms. Approximately 2.4% of queries took longer than 300 ms, about 1.5% exceeded 500 ms, and fewer than 1% exceeded 1 s. Outlier slow responses were generally associated with queries requiring higher edit distances or yielding many matching results.

### 3.3.3 Computational and energy footprint

The mean CPU utilisation was 0.358% (median 0.265%, IQR 0.174), corresponding to an equivalent full-load CPU time of approximately 30.9 h over one year. The distribution was strongly right-skewed (skewness 17.55; excess kurtosis 352.39), with a 99th percentile spike threshold of 1.400%. Monthly CPU behaviour varied significantly (Kruskal–Wallis $H = 764.34$, $p < 0.001$) (Figure 6); chronological quartiles also differed significantly ($H = 354.15$, $p < 0.001$), with median utilisation increasing from Q1 (0.219%) to Q4 (0.292%). The heatmap demonstrated a reproducible diurnal pattern, with lower utilisation during early morning hours and higher utilisation during late morning to afternoon periods.

TABLE 1 Top 20 individual medicines searched on MediVerify and their listings on the WHO Model EML (2025) and Sri Lanka National EML (2022). This divergence has direct implications for EML revision and for understanding the evolving disease burden and prescribing landscape in Sri Lanka.

| Rank | Medicine | WHO EML 2025 | Sri Lanka EML 2022 |
|---|---|---|---|
| 1 | Rosuvastatin | No | No |
| 2 | Paracetamol | Yes | No |
| 3 | Glimepiride | No | No |
| 4 | Montelukast | No | No |
| 5 | Ondansetron | Yes | Yes |
| 6 | Vitamin E (tocopherol) | No | No |
| 7 | Risperidone | Yes | Yes |
| 8 | Betamethasone | Yes | No |
| 9 | Vitamin B preparations | No | No |
| 10 | Olanzapine | Yes | Yes |
| 11 | Amlodipine | Yes | No |
| 12 | Bisoprolol | No | No |
| 13 | Vitamin C (ascorbic acid) | No | No |
| 14 | Diclofenac | No | No |
| 15 | Tacrolimus | Yes | Yes |
| 16 | Povidone-iodine | Yes | No |
| 17 | Methylprednisolone | No | Yes |
| 18 | Silver sulfadiazine | Yes | Yes |
| 19 | Insulin (human) | Yes | Yes |
| 20 | Calcium supplements | No | No |

Estimated annual energy consumption: ~6.49 kWh, equating to ~0.12 kWh per million queries. Even under a high grid emission factor (0.6–0.7 kg $CO_2$/kWh), this translates to only ~0.06–0.14 kg $CO_2$ per million queries, indicating an extremely low environmental footprint. These estimates are based on simplified power assumptions and represent order-of-magnitude approximations.

## 4 Discussion

This year-long analysis of Sri Lanka's MediVerify platform demonstrates how large-scale medicine search data can provide actionable insights for public health and policy. The findings reveal strong public interest in chronic disease medications, antibiotics, and supplements, broadly aligning with patterns observed at other national and international medicines information platforms.

The dominance of vitamins/minerals and antibiotics in search behaviour mirrors findings from NHS Open Prescribing in England, where antibiotic prescriptions account for a disproportionate share of primary care consultations and where public interest in supplement-drug interactions is well documented (3). The high search volume for rosuvastatin, glimepiride, and amlodipine is consistent with Moon et al., who found these drug classes frequently prescribed across LMICs regardless of EML status (7). The parallel between Sri Lanka's MediVerify data and these international findings suggests that the country's NCD burden is driving information-seeking in a manner analogous to higher-income countries, despite substantial differences in healthcare system architecture.

The divergence between frequently searched medicines and EML designations aligns with Nkosi et al., who observed substantial public demand for medicines outside formal essential medicines frameworks across sub-Saharan African LMIC settings (8, 15). Population-level search analytics may thus serve as a near-real-time proxy for evolving public demand and prescribing practice, complementing the episodic expert consensus processes that currently underpin EML updates. This represents a novel application of digital health data to inform pharmaceutical policy, with potential for replication across other LMIC regulatory settings.

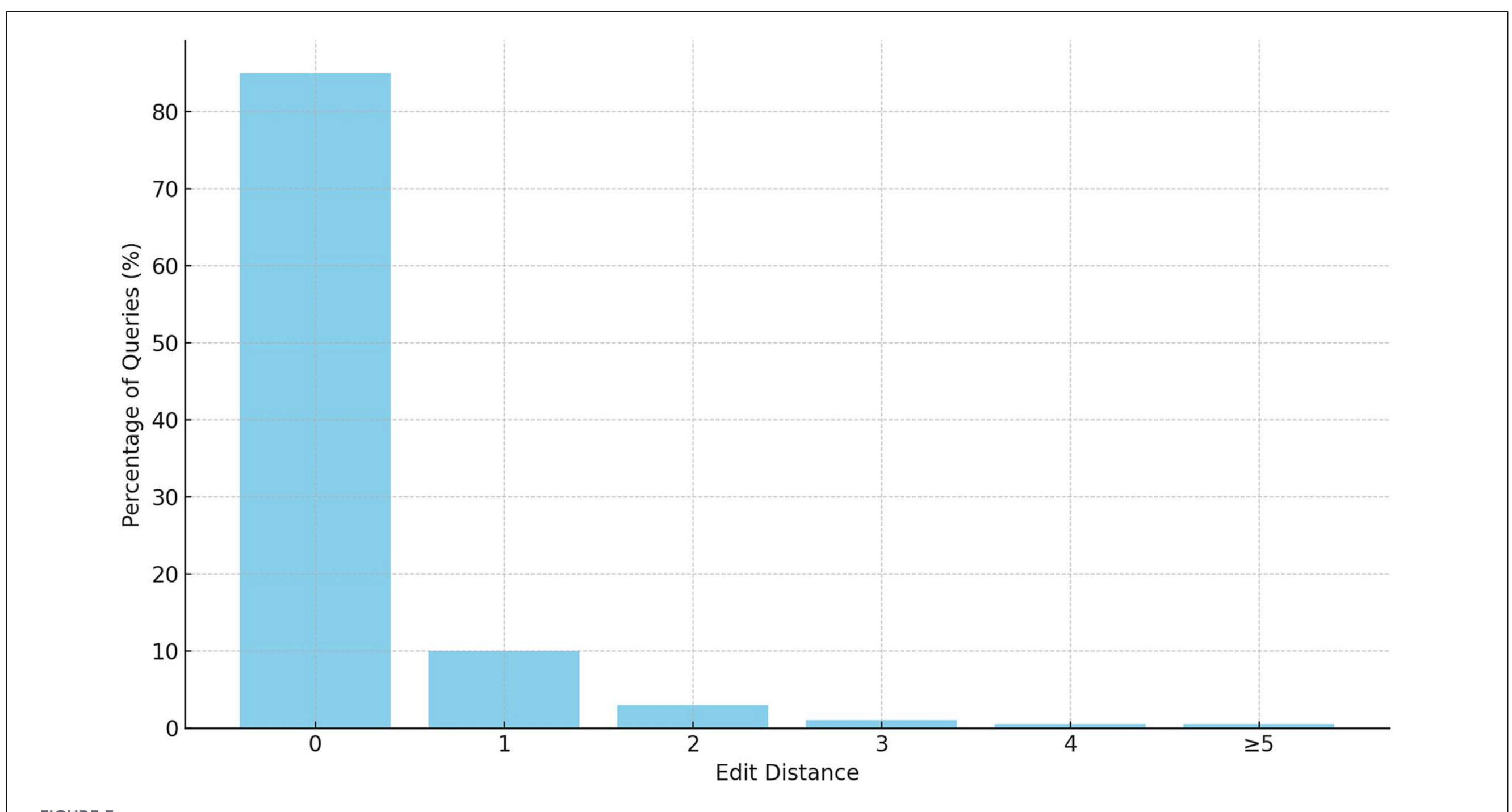

FIGURE 3
Distribution of string edit distances for MediVerify searches over one year. An edit distance of 0 indicates an exact match. The high proportion of exact matches (≈85%) indicates that users generally enter medicine names accurately, suggesting adequate health and digital literacy for self-directed medicine information seeking among MediVerify users.

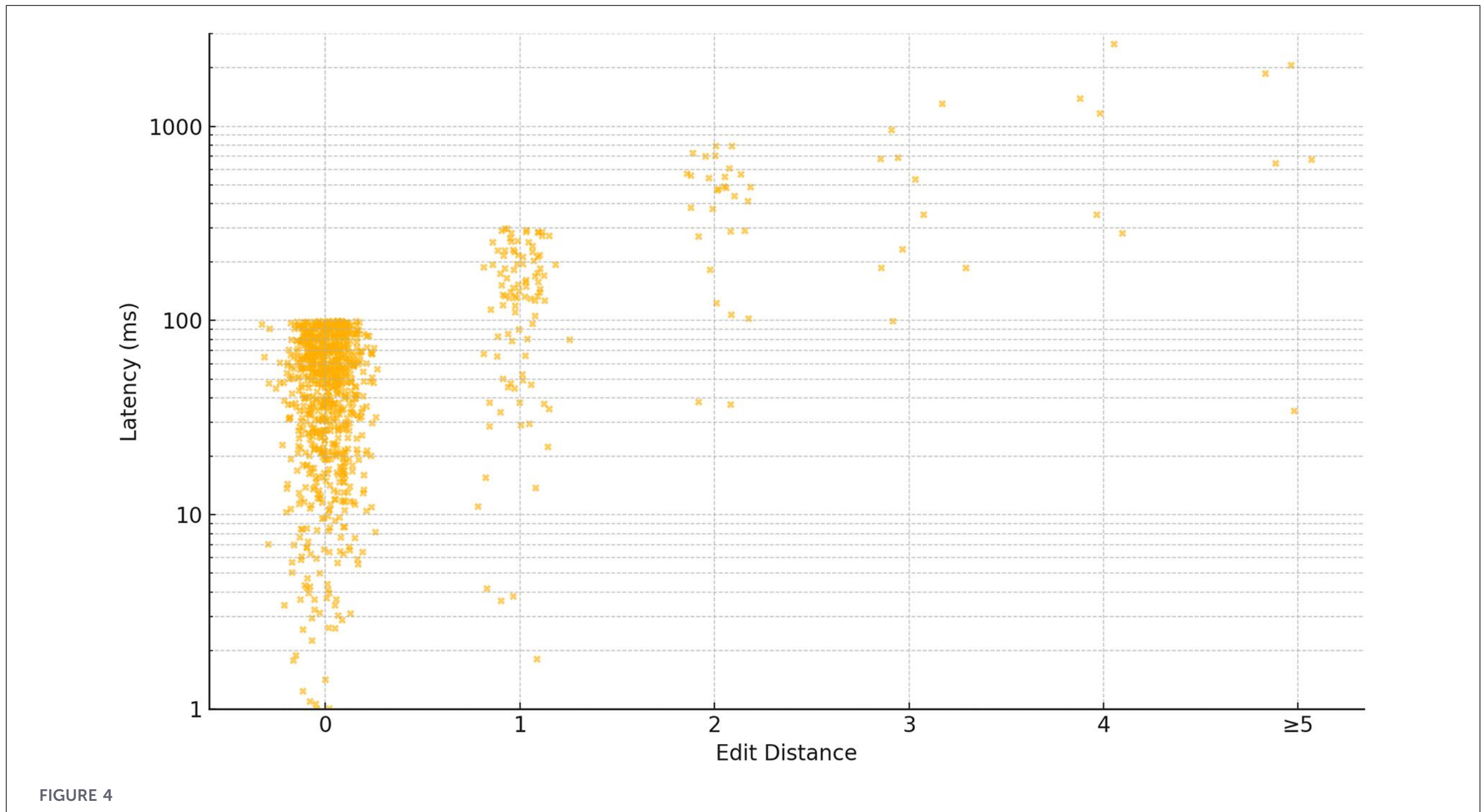

FIGURE 4
Histogram of query response times (latency) on the MediVerify platform, plotted on a logarithmic scale. Over 90% of queries completed in under 50 ms. The practical implication is that MediVerify delivers a near-instantaneous user experience for the overwhelming majority of users—a critical feature for a national platform serving a population with varying digital literacy and internet connection speeds.

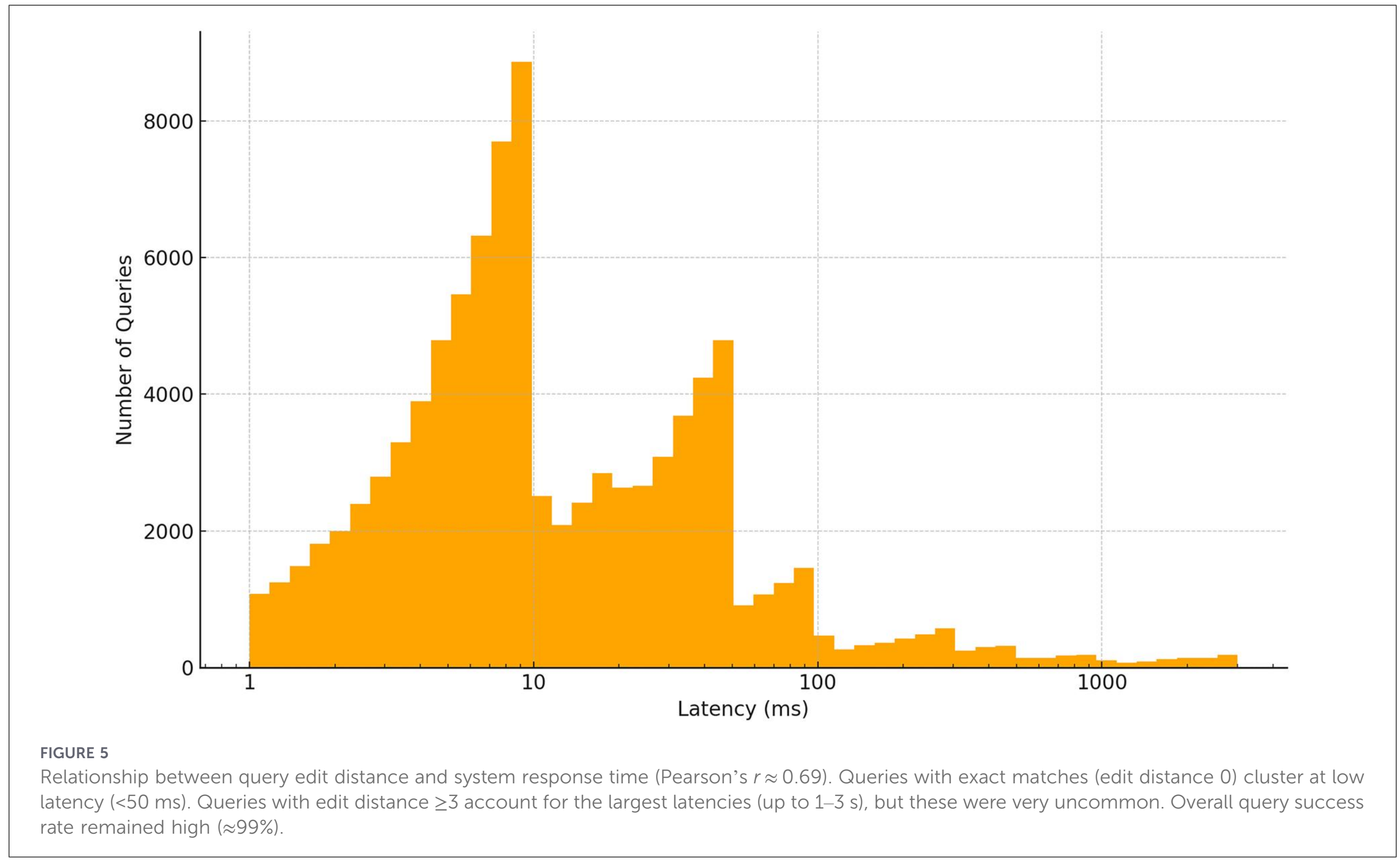

FIGURE 5
Relationship between query edit distance and system response time (Pearson's $r \approx 0.69$). Queries with exact matches (edit distance 0) cluster at low latency (<50 ms). Queries with edit distance ≥3 account for the largest latencies (up to 1–3 s), but these were very uncommon. Overall query success rate remained high (≈99%).

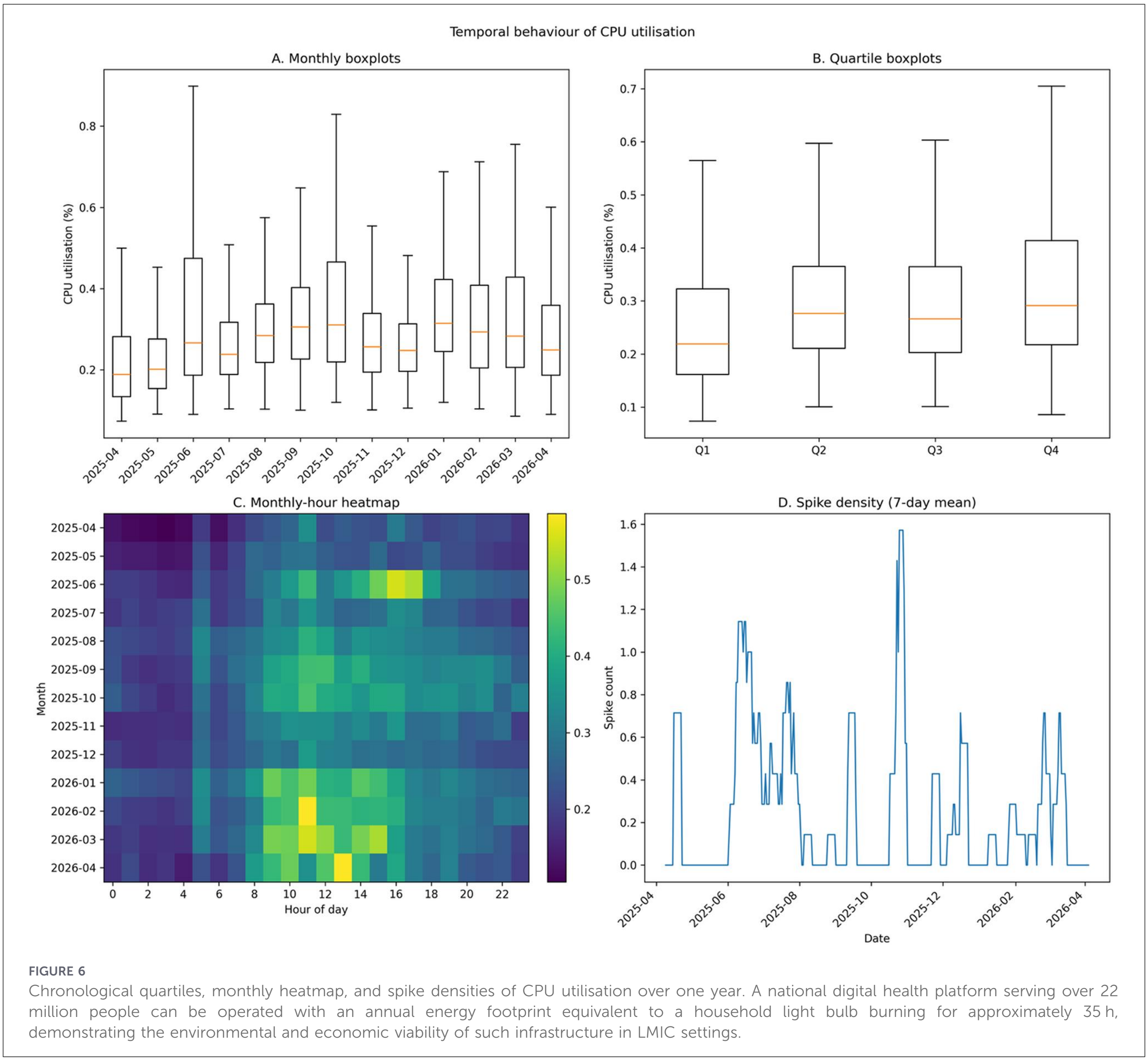

FIGURE 6
Chronological quartiles, monthly heatmap, and spike densities of CPU utilisation over one year. A national digital health platform serving over 22 million people can be operated with an annual energy footprint equivalent to a household light bulb burning for approximately 35 h, demonstrating the environmental and economic viability of such infrastructure in LMIC settings.

Analysis of zero-result queries provided further insight into unmet information needs. Although only ~1% of all queries, the systematic monitoring of these queries could support targeted database improvements, including incorporation of synonyms, spelling variants, and registration of in-demand products such as mebendazole. The high proportion of zero-query medicines (40%–50%) is consistent with a “long tail” pattern also reported for NHS Open Prescribing data in England, where a small subset of drugs accounts for the vast majority of national prescribing volume.

From a technical perspective, MediVerify demonstrated robust performance at national scale. The fuzzy matching implementation (Levenshtein distance, threshold ≤5) maintained acceptable performance even for difficult queries, with nearly all searches completing in under one second. The extremely low energy footprint (~0.12 kWh per million queries) confirms that such platforms can be deployed sustainably in resource-constrained environments. While energy estimates are based on simplified assumptions, they confirm the negligible resource requirements associated with well-designed search infrastructure.

# 5 Conclusions

## 5.1 National medicine information-seeking behaviour

MediVerify data reveal that public information-seeking in Sri Lanka is concentrated in a small number of therapeutic categories reflecting the country’s NCD burden, with strong temporal variation across quarters providing actionable data for platform maintenance and public health communication scheduling. Zero-result query monitoring is a sensitive tool for identifying registry gaps that can be systematically addressed through database updates and public education.

## 5.2 Essential medicines policy alignment

A significant divergence exists between the medicines most frequently searched by the Sri Lankan public and those designated as essential at national and international levels, with ~70% of the top 20 searched medicines absent from one or both EMLs. This likely reflects evolving disease burdens and changing prescribing practices in a setting undergoing epidemiological transition. Population-level search analytics offer a novel, near-real-time data source for informing periodic EML revision and should be formally integrated into the review cycle.

## 5.3 System performance and sustainability

MediVerify demonstrates that a national digital health platform can achieve high utilisation (>1.49 million queries in year one) with robust performance (median latency 8 ms, >99% query resolution) and minimal computational and environmental overhead (~0.12 kWh per million queries). These characteristics confirm the viability of replicating this model across other LMIC national regulatory authorities, where resource constraints may otherwise inhibit digital health investment.

## 5.4 Policy recommendations from this study

### 5.4.1 National medicines regulatory authority (NMRA)

(1) Integrate synonym and brand-name variants into the MediVerify search index to reduce zero-result queries from brand-generic mismatches. (2) Address identified registry gaps, including the registration of mebendazole and other high-frequency zero-result medicines. (3) Implement systematic monthly review of zero-result query logs as a pharmacovigilance and registry maintenance tool. (4) Expand the NMRA medicines database to include patient information leaflets and regulatory approval status history.

### 5.4.2 Ministry of health and nutrition

(1) Commission a formal review of the Sri Lanka National EML informed by MediVerify demand data, specifically regarding rosuvastatin, glimepiride, bisoprolol, montelukast, and other highly searched non-essential medicines. (2) Integrate MediVerify query analytics into the national Pharmaceutical Management Information System (PMIS) as a demand-side data source. (3) Share the MediVerify model and methodology with the WHO SEARO regional network to facilitate adoption by other LMIC national regulatory authorities in the region.

### 5.4.3 Provincial and sub-national health directorates

(1) Develop targeted public communication campaigns to raise awareness of essential medicines in the "zero-query" category, particularly life-saving medicines where public under-awareness is a concern. (2) Incorporate MediVerify analytics into provincial health planning cycles to identify local variations in medicine information demand. (3) Provide pharmacist and community health worker training on MediVerify as a first-line medicines information resource, reducing reliance on unverified online sources.

# 6 Limitations

This study has several limitations. First, search queries may not directly reflect actual medicine utilisation or prescribing patterns. Second, the absence of user-level demographic data limits interpretation of population subgroups. Third, registry completeness and naming conventions may influence zero-result and zero-query findings. Fourth, energy consumption estimates are based on simplified assumptions and represent order-of-magnitude approximations rather than precise measurements. Fifth, the study is limited to a single national platform in a single LMIC context; generalisability to other national regulatory settings requires validation.

# Data availability statement

The raw data supporting the conclusions of this article will be made available by the authors, without undue reservation.

# Ethics statement

Ethical approval was not required for the study involving humans in accordance with the local legislation and institutional requirements. Written informed consent to participate in this study was not required from the participants or the participants' legal guardians/next of kin in accordance with the national legislation and the institutional requirements.

# Author contributions

PCA: Conceptualization, Data curation, Formal analysis, Investigation, Software, Validation, Visualization, Writing – original draft, Writing – review & editing. PMA: Conceptualization, Data curation, Formal analysis, Investigation, Resources, Supervision, Writing – original draft, Writing – review & editing. RF: Conceptualization, Data curation, Project administration, Resources, Supervision, Writing – original draft, Writing – review & editing.

# Funding

The author(s) declared that financial support was not received for this work and/or its publication.

# Conflict of interest

The author(s) declared that this work was conducted in the absence of any commercial or financial relationships that could be construed as a potential conflict of interest.

## Generative AI statement

The author(s) declared that generative AI was used in the creation of this manuscript. The authors verify and take full responsibility for the use of generative AI in the preparation of this manuscript. ChatGPT version 5 was used for image generation and linguistic review.

Any alternative text (alt text) provided alongside figures in this article has been generated by Frontiers with the support of artificial intelligence and reasonable efforts have been made to ensure accuracy, including review by the authors wherever possible. If you identify any issues, please contact us.

## Publisher's note

All claims expressed in this article are solely those of the authors and do not necessarily represent those of their affiliated organizations, or those of the publisher, the editors and the reviewers. Any product that may be evaluated in this article, or claim that may be made by its manufacturer, is not guaranteed or endorsed by the publisher.